\documentclass[11pt]{article}
\usepackage{txfonts}
\usepackage{geometry}
\usepackage[utf8]{inputenc}
\usepackage{enumitem,amssymb}
\usepackage{ragged2e}
\newlist{thematic}{itemize}{8}
\setlist[thematic]{label=$\square$}
\usepackage{pifont}
\usepackage{setspace}
\usepackage{eso-pic,graphicx}
\usepackage{tikz}
\usepackage{atbegshi}
\usepackage{url}
\usepackage{xcolor}    
\usepackage{multicol}
\usepackage{fontawesome}
\usepackage[english]{babel}
\usepackage[numbers]{natbib}
\usepackage{authblk}
\usepackage[font=small,labelfont=bf]{caption}
\definecolor{myblue}{RGB}{0,121,194}
\usepackage[colorlinks=true,
            linkcolor=myblue,
            urlcolor=myblue,
            citecolor=myblue]{hyperref}

\begin{document}

\selectlanguage{english}

\raggedright
\Large
ESO Expanding Horizons  \linebreak
\small
Transforming Astronomy in the 2040s \linebreak
Call for White Papers

\vspace{2.cm}
\begin{spacing}{1.6}
\textbf{\fontsize{22pt}{40pt}\selectfont
Probing General Relativity on Cosmological Scales in the 2040s}
\end{spacing}
\normalsize
\vspace{0.5cm}

\textbf{Authors:} Federico Montano$^{1,2}$, Samantha J. Rossiter$^{1,2}$, Chris Addis$^{3}$, Jessie Hammond$^{3}$, Stefano Camera$^{1,2,4}$, Chris Clarkson$^{3,5}$, Mohamed Yousry Elkhashab$^{6,7,8,9}$, Massimo Guidi$^{10,11}$, Ofer Lahav$^{12}$,
Giovanni Aric\`o$^{13}$, Sofia Contarini$^{14}$, Pratika Dayal$^{15}$, Giulia Degni$^{16}$, Antonio Farina$^{17}$, Vid Ir\v{s}i\v{c}$^{18}$, Federico Marulli$^{10}$, Elena Sarpa$^{7}$, Simone Sartori$^{16}$, Emiliano Sefusatti$^{7}$, Francesco Verdiani$^{19}$, and Giovanni Verza$^{20}$

\vspace{0.3cm}

\textbf{Contacts:} \href{federico.montano@unito.it}{federico.montano@unito.it}

\vspace{0.3cm}

\textbf{Affiliations:}\\
{\footnotesize
$^{1}$ Dipartimento di Fisica, Universit\`a degli Studi di Torino, Via P.\ Giuria 1, 10125 Torino, Italy \\
$^{2}$ INFN -- Istituto Nazionale di Fisica Nucleare, Sezione di Torino, Via P.\ Giuria 1, 10125 Torino, Italy \\
$^{3}$ Astronomy Unit, School of Physical \& Chemical Sciences, Queen Mary University of London, London E1 4NS, UK\\
$^{4}$ INAF -- Istituto Nazionale di Astrofisica, Osservatorio Astrofisico di Torino, Strada Osservatorio 20, 10025 Pino Torinese, Italy\\
$^{5}$ Department of Physics \& Astronomy, University of Western Cape, Cape Town 7535, South Africa \\
$^{6}$ Dipartimento di Fisica – Sezione di Astronomia, Universit\`a di Trieste, Via Tiepolo 11, 34131 Trieste, Italy \\
$^{7}$ INAF -- Osservatorio Astronomico di Trieste, Via G.\ B.\ Tiepolo 11, 34131 Trieste, Italy \\
$^{8}$ INFN -- Sezione di Trieste, Via Valerio 2, 34127 Trieste TS, Italy \\
$^{9}$ IFPU, Institute for Fundamental Physics of the Universe, Via Beirut 2, 34151 Trieste, Italy \\
$^{10}$ Dipartimento di Fisica e Astronomia ``Augusto Righi'', Alma Mater Studiorum Universit\`a di Bologna, 40129 Bologna, Italy\\
$^{11}$ INAF -- Osservatorio di Astrofisica e Scienza dello Spazio di Bologna, Via P.\ Gobetti 93/3, 40129 Bologna, Italy\\
$^{12}$ Department of Physics and Astronomy, University College London, Gower Street, London WC1E 6BT, UK\\
$^{13}$ INFN -- Sezione di Bologna, Viale Berti Pichat 6/2, 40127 Bologna, Italy\\
$^{14}$ Max Planck Institute for Extraterrestrial Physics, Giessenbachstrasse 1, 85748 Garching, Germany\\
$^{15}$ Canadian Institute for Theoretical Astrophysics, University of Toronto, Toronto, ON, Canada\\
$^{16}$ Aix Marseille Universit\'e, CNRS/IN2P3, CPPM, Marseille, France\\
$^{17}$ INAF -- Osservatorio Astronomico di Brera, Via Brera 28, 20122 Milano, Italy \\
$^{18}$ Center for Astrophysics Research, Department of Physics, Astronomy and Mathematics, University of Hertfordshire, Hatfield AL10 9AB, UK\\
$^{19}$ Scuola Internazionale Superiore di Studi Avanzati (SISSA), Via Bonomea 265, 34136 Trieste,
Italy\\
$^{20}$ ICTP -- International Centre for Theoretical Physics, Strada Costiera 11, 34151, Trieste, Italy\\
}

\pagenumbering{gobble} 

\pagebreak

\justifying

\section*{Relativistic effects in the large-scale structure of the universe}

The effectiveness of our description of the Universe's evolution mostly lies in our understanding of its fundamental driver, the gravitational interaction. The theory of general relativity (GR)  has been confirmed with exquisite accuracy in strong field regimes. However, on cosmological scales, where gravity governs structure formation, direct observational tests remain comparatively weak. 

Large-scale structure (LSS) surveys offer a unique probe of GR on these ultra-large scales, which are intrinsically limited by cosmic variance---the low statistical sampling of long-wavelength modes available in a given survey volume. Relativistic effects perturb the observed redshift and angular position of galaxies, imprinting distortions in both the radial and transverse clustering pattern. Small corrections to redshift-space galaxy number counts arise because observations lie on the past light cone, where metric perturbations modify photon paths.  These contributions represent ``smoking-gun" signatures of GR, and gain significance as scales approach the horizon. 
As a result, detection of relativistic corrections requires vast survey volumes, which upcoming galaxy surveys promise to deliver. 

Perturbation theory provides concrete predictions for the amplitude and form of relativistic corrections, which encompass local and integrated contributions \cite{2010Yoo,2011Bonvin&Durrer,2011Challinor&Lewis}. 
The former originate from fields evaluated at the galaxy’s position, such as Doppler shifts, gravitational redshift, and contributions from the local gravitational potential. The latter accumulate along the line of sight and include gravitational lensing magnification, Shapiro time-delay and the integrated Sachs–Wolfe effect. 
Additionally, non-linear evolution generates additional ``dynamical" relativistic terms relevant for higher order statistics \cite{2016VillaRampf}.
Robust modelling of GR effects is a key requirement for upcoming cosmological analyses as it enables direct detection of relativistic contributions in LSS observables and constraints on the evolution of metric perturbations as a function of redshift, uniquely probing spacetime on top of its background description.

Different relativistic terms dominate at different epochs \cite{2022Castorina&DiDio}. Doppler contributions, given by the peculiar velocity of sources, are dominant at low redshift, while potential terms and integrated corrections such as lensing gain relevance as redshift increases. While velocity contributions will likely be detectable already with upcoming (Stage IV) low-$z$ data, it would be a transformative novelty to measure the pure GR signal coming from the metric potentials themselves. However, wide, high-redshift spectroscopic cosmological surveys are needed to achieve this.

Crucially, not only are GR corrections a potential signal, but also a systematic. As survey volumes grow and statistical uncertainties shrink, GR effects can mimic primordial non-Gaussianity and bias cosmological parameters \cite{2015Camera_,2025Baldi_}. Incorporating them into theoretical predictions and analysis pipelines is thus essential for both unbiased parameter estimation and testing GR or modified gravity scenarios. 
Such tests shall concern, for instance, the validity of the equivalence principle \cite{2018Bonvin&Fleury,2024Castello_}, the presence of gravitational slip \cite{2007Zhang_}, or the determination of our peculiar velocity as inferred from the finger-of-the-observer signal \cite{2025Elkhashab_}.

ESO's Expanding Horizon facilities will enable this scientific leap. By mapping the high-redshift universe with unprecedented precision and depth, a new window will open to GR in the ultra-large scale, deeply linear regime, inaugurating an era of high-redshift precision cosmology. 

\section*{Observational status}
\begin{figure}
    \centering
    \includegraphics[width=0.8\linewidth]{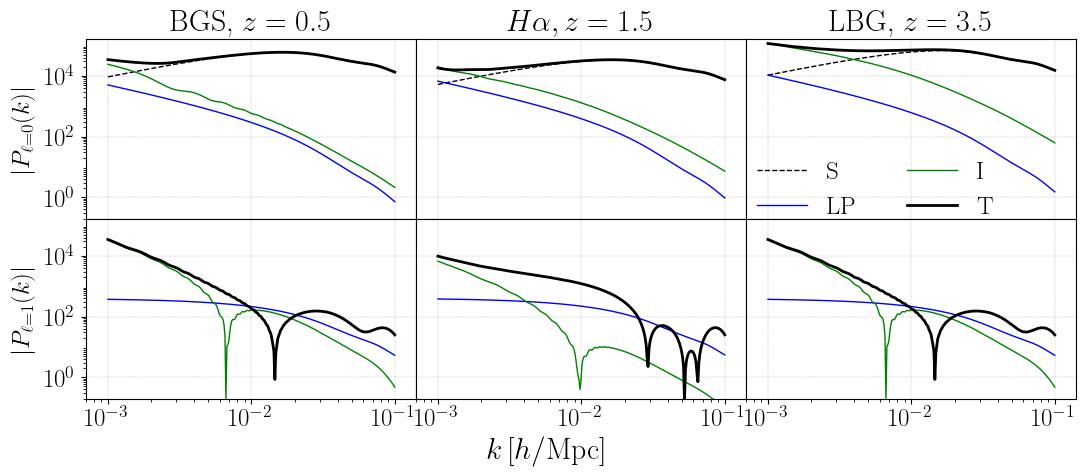}
    \caption{Monopole (top) of the 3D power spectrum for a bright galaxy sample (BGS, $m_{\rm c}=20$), an H$\alpha$ ($F_{\rm c} = 2 \times 10^{-16} \, {\rm erg\,cm^{-2}\,s^{-1}}$) and a Lyman-break galaxy (LBG, $m_{\rm c} = 25$) survey and the dipole contribution (bottom) for a bright-faint split cross-power spectrum, in units of $(\mathrm{Mpc}/h)^{3}$.
    Evaluated using \href{https://github.com/craddis1/CosmoWAP}{\faGithub},
    the total signal (T) is decomposed into its main components: standard (S, i.e.\ Newtonian), local GR projection effects (LP), and integrated GR terms (I). 
    }
    \label{fig:PkPlot}
\end{figure}

To date, observational efforts to detect GR corrections in galaxy clustering measurements remain inconclusive, exceptions being gravitational redshift in galaxy clusters \cite{2011Wojtak_,2015Sadeh_} and lensing magnification \cite{1998Moessner_,2023Elvin_Poole_}. No definitive measurements have yet been reported with major Stage-III surveys, nor with early Stage-IV data, like DESI \cite{2024DESI_} or \textit{Euclid} \cite{2025Euclid}. Analyses of two- and three-point statistics of the galaxy distribution have tightly constrained redshift-space distortions but lack the sensitivity required to isolate peculiar relativistic signals on cosmological scales, like Doppler or gravitational redshift. These contributions manifest in the imaginary, parity-odd, components of the power spectrum \cite{2009McDonald}. Final Stage-IV data are expected to deliver the first meaningful constraints on these asymmetric GR contributions via cross-correlation measurements. 
Importantly, the prospects for detecting relativistic contributions exhibit a clear redshift dependence (as Fig.\ \ref{fig:PkPlot} depicts), driven by the scales and tracer densities accessible to different surveys. 

At low redshift, dense spectroscopic samples provide optimal conditions for isolating the dipole generated by velocity terms. These Doppler-induced asymmetries represent the most promising first-detection channel.
At high redshift, forthcoming Stage-V surveys will sample enormous cosmic volumes, offering the first opportunity to probe relativistic corrections dominated by contributions from metric potentials \cite{2025Addis_}. The underlying dark matter distribution at these epochs is traced by Lyman-break galaxies (LBGs), whose large number densities and high bias amplify sensitivity to line-of-sight GR effects. Probing these signatures will provide a new test of gravity, directly linking the evolution of spacetime perturbations to the growth of structures. Detection of relativistic effects thus places stringent demands on instrumentation and survey design.

\section*{Technology requirements}
LBGs are high-redshift ($2 \lesssim z \lesssim 5$) star-forming galaxies identified through a characteristic spectral drop in their ultraviolet continuum due to absorption by neutral hydrogen. This selection technique enables efficient identification of large samples of galaxies across cosmic time. 
Wide-field of view spectroscopic surveys such as the WST \cite{2024WST_} or MegaMapper \cite{2022MegaMapper_} shall aim to observe large patches of the sky ($\gtrsim 10\,000 \,\deg^2$)  with unprecedented sensitivity and multiplexing ($\,>20\,000$ fibres), reaching an LBG target density $\gtrsim 10^{-4} \, h^3\, {\rm Mpc}^{-3}$. This will empower us with the statistics required to constrain large-scale effects even at high-$z$, with galaxy clustering data alone and in combination with gravitational wave sources \cite{2025Zazzera_}. 

\begin{figure}
    \centering
    \includegraphics[width=0.48\linewidth]{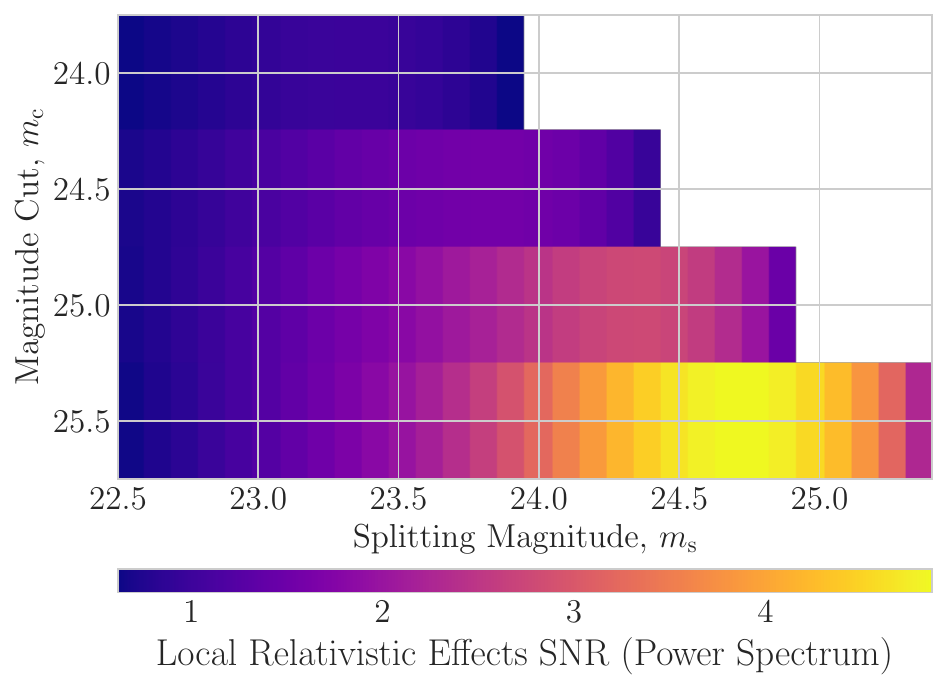}
    \includegraphics[width=0.48\linewidth]{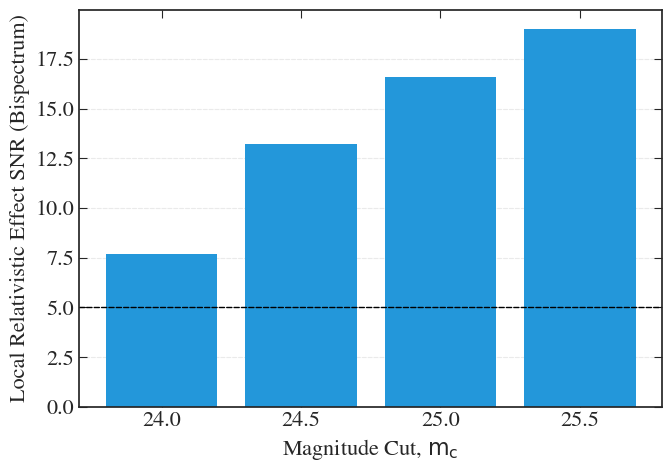}
    \caption{Estimate of SNR of local relativistic effects in a LBG population (modelled as in \cite{2019Wilson&White}, in $z\in[2,\,5]$), assuming a WST-like \cite{2024WST_} redshift survey, with a conservative sky coverage of $10\,00 \;\rm deg^2$. \emph{Left:} Colourmap of the SNR in a faint-bright multi-tracer full-power spectrum measurement, $P(k,\,\mu),$ with $\mu\in[0,\,1]$, as a function of magnitude cut and split \cite{2024Montano&Camera}. \emph{Right:} Total SNR in an auto-bispectrum analysis; deeper cuts yield higher significance \cite{2025Rossiter_}.
    }
    \label{fig:SNR}
\end{figure}

Local relativistic effects, dominated by the Doppler term, can be isolated thanks to a multi-tracer approach, which enables us to jointly analyse auto- and cross-power spectra \cite{2009Seljak}, thereby reducing cosmic variance. By splitting the LBG population into a faint and bright sub-sample \cite{2014Bonvin_}, a detection is feasible already with a conservative surveyed area of $10\,000 \,\deg^2$, as shown in Fig.\ \ref{fig:SNR} (left). 
Furthermore, Fig.\ \ref{fig:SNR} (right) shows that ${>}5\,\sigma$ detections are possible using the single-tracer bispectrum, which retains the odd-parity, leading order relativistic signature.
In both cases, estimates of signal-to-noise ratio (SNR) of local relativistic effects are reported for different magnitude cuts. Since the SNR scales with
the square root of the observed sky fraction
, increases in the planned survey coverage will improve the likelihood of a significant detection. 

Together, these requirements delineate a clear technological pathway toward transforming high-redshift galaxy surveys into laboratories for relativistic cosmology. By pushing spectroscopic mapping to unprecedented depth, area, and multiplexing, 2040s facilities will turn relativistic signatures from marginal detections into precision observables, thereby extending LSS analyses beyond the Newtonian regime.

\small
\bibliographystyle{apsrev4-1}
\bibliography{sample}

@ARTICLE{2016VillaRampf,
       author = {{Villa}, Eleonora and {Rampf}, Cornelius},
        title = "{Relativistic perturbations in {\ensuremath{\Lambda}}CDM: Eulerian \& Lagrangian approaches}",
      journal = {\jcap},
         year = 2016,
        month = jan,
       volume = {2016},
       number = {1},
        pages = {030-030},
          doi = {10.1088/1475-7516/2016/01/030},
archivePrefix = {arXiv},
       eprint = {1505.04782},
 primaryClass = {gr-qc},
       adsurl = {https://ui.adsabs.harvard.edu/abs/2016JCAP...01..030V}
}

@ARTICLE{2024WST_,
       author = {{Mainieri}, Vincenzo and others},
        title = "{The Wide-field Spectroscopic Telescope (WST) Science White Paper}",
      journal = {arXiv e-prints},
         year = 2024,
        month = mar,
          eid = {arXiv:2403.05398},
        pages = {arXiv:2403.05398},
          doi = {10.48550/arXiv.2403.05398},
archivePrefix = {arXiv},
       eprint = {2403.05398},
 primaryClass = {astro-ph.IM},
       adsurl = {https://ui.adsabs.harvard.edu/abs/2024arXiv240305398M}
}

@ARTICLE{2019Wilson&White,
       author = {{Wilson}, M.~J. and {White}, Martin},
        title = "{Cosmology with dropout selection: straw-man surveys \& CMB lensing}",
      journal = {\jcap},
         year = 2019,
        month = oct,
       volume = {2019},
       number = {10},
          eid = {015},
        pages = {015},
          doi = {10.1088/1475-7516/2019/10/015},
archivePrefix = {arXiv},
       eprint = {1904.13378},
 primaryClass = {astro-ph.CO},
       adsurl = {https://ui.adsabs.harvard.edu/abs/2019JCAP...10..015W}
}

@ARTICLE{2011Bonvin&Durrer,
       author = {{Bonvin}, Camille and {Durrer}, Ruth},
        title = "{What galaxy surveys really measure}",
      journal = {\prd},
         year = 2011,
        month = sep,
       volume = {84},
       number = {6},
          eid = {063505},
        pages = {063505},
          doi = {10.1103/PhysRevD.84.063505},
archivePrefix = {arXiv},
       eprint = {1105.5280},
 primaryClass = {astro-ph.CO},
       adsurl = {https://ui.adsabs.harvard.edu/abs/2011PhRvD..84f3505B}
}

@ARTICLE{2011Challinor&Lewis,
       author = {{Challinor}, Anthony and {Lewis}, Antony},
        title = "{Linear power spectrum of observed source number counts}",
      journal = {\prd},
         year = 2011,
        month = aug,
       volume = {84},
       number = {4},
          eid = {043516},
        pages = {043516},
          doi = {10.1103/PhysRevD.84.043516},
archivePrefix = {arXiv},
       eprint = {1105.5292},
 primaryClass = {astro-ph.CO},
       adsurl = {https://ui.adsabs.harvard.edu/abs/2011PhRvD..84d3516C}
}

@ARTICLE{2010Yoo,
       author = {{Yoo}, Jaiyul},
        title = "{General relativistic description of the observed galaxy power spectrum: Do we understand what we measure?}",
      journal = {\prd},
         year = 2010,
        month = oct,
       volume = {82},
       number = {8},
          eid = {083508},
        pages = {083508},
          doi = {10.1103/PhysRevD.82.083508},
archivePrefix = {arXiv},
       eprint = {1009.3021},
 primaryClass = {astro-ph.CO},
       adsurl = {https://ui.adsabs.harvard.edu/abs/2010PhRvD..82h3508Y}
}

@ARTICLE{2009McDonald,
       author = {{McDonald}, Patrick},
        title = "{Gravitational redshift and other redshift-space distortions of the imaginary part of the power spectrum}",
      journal = {\jcap},
         year = 2009,
        month = nov,
       volume = {2009},
       number = {11},
          eid = {026},
        pages = {026},
          doi = {10.1088/1475-7516/2009/11/026},
archivePrefix = {arXiv},
       eprint = {0907.5220},
 primaryClass = {astro-ph.CO},
       adsurl = {https://ui.adsabs.harvard.edu/abs/2009JCAP...11..026M}
}

@ARTICLE{2014Bonvin_,
       author = {{Bonvin}, Camille and {Hui}, Lam and {Gazta{\~n}aga}, Enrique},
        title = "{Asymmetric galaxy correlation functions}",
      journal = {\prd},
         year = 2014,
        month = apr,
       volume = {89},
       number = {8},
          eid = {083535},
        pages = {083535},
          doi = {10.1103/PhysRevD.89.083535},
archivePrefix = {arXiv},
       eprint = {1309.1321},
 primaryClass = {astro-ph.CO},
       adsurl = {https://ui.adsabs.harvard.edu/abs/2014PhRvD..89h3535B}
}

@ARTICLE{2024Montano&Camera,
       author = {{Montano}, Federico and {Camera}, Stefano},
        title = "{Detecting relativistic Doppler by multi-tracing a single galaxy population}",
      journal = {Physics of the Dark Universe},
         year = 2024,
        month = dec,
       volume = {46},
          eid = {101634},
        pages = {101634},
          doi = {10.1016/j.dark.2024.101634},
archivePrefix = {arXiv},
       eprint = {2407.06284},
 primaryClass = {astro-ph.CO},
       adsurl = {https://ui.adsabs.harvard.edu/abs/2024PDU....4601634M}
}

@ARTICLE{2025Addis_,
       author = {{Addis}, Chris and others},
        title = "{Unbiased analysis of primordial non-Gaussianity: the multipoles of the full relativistic power spectrum}",
      journal = {arXiv e-prints},
         year = 2025,
        month = nov,
          eid = {arXiv:2511.09466},
        pages = {arXiv:2511.09466},
          doi = {10.48550/arXiv.2511.09466},
archivePrefix = {arXiv},
       eprint = {2511.09466},
 primaryClass = {astro-ph.CO},
       adsurl = {https://ui.adsabs.harvard.edu/abs/2025arXiv251109466A}
}

@ARTICLE{2025Rossiter_,
       author = {{Rossiter}, Samantha J. and others},
        title = "{Decoupling local primordial non-Gaussianity from relativistic effects in the galaxy bispectrum}",
      journal = {\jcap},
         year = 2025,
        month = jul,
       volume = {2025},
       number = {7},
          eid = {055},
        pages = {055},
          doi = {10.1088/1475-7516/2025/07/055},
archivePrefix = {arXiv},
       eprint = {2407.06301},
 primaryClass = {astro-ph.CO},
       adsurl = {https://ui.adsabs.harvard.edu/abs/2025JCAP...07..055R}
}

@ARTICLE{2025Euclid,
       author = {{Euclid Collaboration} and {Mellier}, Y. and others},
        title = "{Euclid: I. Overview of the Euclid mission}",
      journal = {\aap},
         year = 2025,
        month = may,
       volume = {697},
          eid = {A1},
        pages = {A1},
          doi = {10.1051/0004-6361/202450810},
archivePrefix = {arXiv},
       eprint = {2405.13491},
 primaryClass = {astro-ph.CO},
       adsurl = {https://ui.adsabs.harvard.edu/abs/2025A&A...697A...1E}
}

@ARTICLE{2024DESI_,
       author = {{DESI Collaboration} and {Adame}, A.~G. and others},
        title = "{Validation of the Scientific Program for the Dark Energy Spectroscopic Instrument}",
      journal = {\aj},
         year = 2024,
        month = feb,
       volume = {167},
       number = {2},
          eid = {62},
        pages = {62},
          doi = {10.3847/1538-3881/ad0b08},
archivePrefix = {arXiv},
       eprint = {2306.06307},
 primaryClass = {astro-ph.CO},
       adsurl = {https://ui.adsabs.harvard.edu/abs/2024AJ....167...62D}
}

@ARTICLE{2009Seljak,
       author = {{Seljak}, Uro{\v{s}}},
        title = "{Extracting Primordial Non-Gaussianity without Cosmic Variance}",
      journal = {\prl},
         year = 2009,
        month = jan,
       volume = {102},
       number = {2},
          eid = {021302},
        pages = {021302},
          doi = {10.1103/PhysRevLett.102.021302},
archivePrefix = {arXiv},
       eprint = {0807.1770},
 primaryClass = {astro-ph},
       adsurl = {https://ui.adsabs.harvard.edu/abs/2009PhRvL.102b1302S}
}

@ARTICLE{2015Camera_,
       author = {{Camera}, S. and {Maartens}, R. and {Santos}, M.~G.},
        title = "{Einstein's legacy in galaxy surveys.}",
      journal = {\mnras},
         year = 2015,
        month = jul,
       volume = {451},
        pages = {L80-L84},
          doi = {10.1093/mnrasl/slv069},
archivePrefix = {arXiv},
       eprint = {1412.4781},
 primaryClass = {astro-ph.CO},
       adsurl = {https://ui.adsabs.harvard.edu/abs/2015MNRAS.451L..80C}
}

@ARTICLE{2024Castello_,
       author = {{Castello}, Sveva and others},
        title = "{Gravitational redshift constraints on the effective theory of interacting dark energy}",
      journal = {\jcap},
         year = 2024,
        month = may,
       volume = {2024},
       number = {5},
          eid = {003},
        pages = {003},
          doi = {10.1088/1475-7516/2024/05/003},
archivePrefix = {arXiv},
       eprint = {2311.14425},
 primaryClass = {astro-ph.CO},
       adsurl = {https://ui.adsabs.harvard.edu/abs/2024JCAP...05..003C}
}

@ARTICLE{2022MegaMapper_,
       author = {{Schlegel}, David J. and others},
        title = "{The MegaMapper: A Stage-5 Spectroscopic Instrument Concept for the Study of Inflation and Dark Energy}",
      journal = {arXiv e-prints},
         year = 2022,
        month = sep,
          eid = {arXiv:2209.04322},
        pages = {arXiv:2209.04322},
          doi = {10.48550/arXiv.2209.04322},
archivePrefix = {arXiv},
       eprint = {2209.04322},
 primaryClass = {astro-ph.IM},
       adsurl = {https://ui.adsabs.harvard.edu/abs/2022arXiv220904322S}
}

@ARTICLE{2022Castorina&DiDio,
       author = {{Castorina}, Emanuele and {Di Dio}, Enea},
        title = "{The observed galaxy power spectrum in General Relativity}",
      journal = {\jcap},
         year = 2022,
        month = jan,
       volume = {2022},
       number = {1},
          eid = {061},
        pages = {061},
          doi = {10.1088/1475-7516/2022/01/061},
archivePrefix = {arXiv},
       eprint = {2106.08857},
 primaryClass = {astro-ph.CO},
       adsurl = {https://ui.adsabs.harvard.edu/abs/2022JCAP...01..061C}
}

@ARTICLE{2025Zazzera_,
       author = {{Zazzera}, Stefano and others},
        title = "{Exploring future synergies for large-scale structure between gravitational waves and radio sources}",
      journal = {arXiv e-prints},
         year = 2025,
        month = may,
          eid = {arXiv:2505.15645},
        pages = {arXiv:2505.15645},
          doi = {10.48550/arXiv.2505.15645},
archivePrefix = {arXiv},
       eprint = {2505.15645},
 primaryClass = {astro-ph.CO},
       adsurl = {https://ui.adsabs.harvard.edu/abs/2025arXiv250515645Z}
}

@ARTICLE{2011Wojtak_,
       author = {{Wojtak}, Rados{\l}aw and {Hansen}, Steen H. and {Hjorth}, Jens},
        title = "{Gravitational redshift of galaxies in clusters as predicted by general relativity}",
      journal = {\nat},
         year = 2011,
        month = sep,
       volume = {477},
       number = {7366},
        pages = {567-569},
          doi = {10.1038/nature10445},
archivePrefix = {arXiv},
       eprint = {1109.6571},
 primaryClass = {astro-ph.CO},
       adsurl = {https://ui.adsabs.harvard.edu/abs/2011Natur.477..567W}
}

@ARTICLE{2015Sadeh_,
       author = {{Sadeh}, Iftach and {Feng}, Low Lerh and {Lahav}, Ofer},
        title = "{Gravitational Redshift of Galaxies in Clusters from the Sloan Digital Sky Survey and the Baryon Oscillation Spectroscopic Survey}",
      journal = {\prl},
         year = 2015,
        month = feb,
       volume = {114},
       number = {7},
          eid = {071103},
        pages = {071103},
          doi = {10.1103/PhysRevLett.114.071103},
archivePrefix = {arXiv},
       eprint = {1410.5262},
 primaryClass = {astro-ph.CO},
       adsurl = {https://ui.adsabs.harvard.edu/abs/2015PhRvL.114g1103S}
}

@ARTICLE{2023Elvin_Poole_,
       author = {{Elvin-Poole}, J. and others},
        title = "{Dark Energy Survey Year 3 results: magnification modelling and impact on cosmological constraints from galaxy clustering and galaxy-galaxy lensing}",
      journal = {\mnras},
         year = 2023,
        month = aug,
       volume = {523},
       number = {3},
        pages = {3649-3670},
          doi = {10.1093/mnras/stad1594},
archivePrefix = {arXiv},
       eprint = {2209.09782},
 primaryClass = {astro-ph.CO},
       adsurl = {https://ui.adsabs.harvard.edu/abs/2023MNRAS.523.3649E}
}

@ARTICLE{1998Moessner_,
       author = {{Moessner}, R. and {Jain}, B. and {Villumsen}, J.~V.},
        title = "{The effect of weak lensing on the angular correlation function of faint galaxies}",
      journal = {\mnras},
         year = 1998,
        month = feb,
       volume = {294},
        pages = {291-298},
          doi = {10.1046/j.1365-8711.1998.01225.x10.1111/j.1365-8711.1998.01225.x},
archivePrefix = {arXiv},
       eprint = {astro-ph/9708271},
 primaryClass = {astro-ph},
       adsurl = {https://ui.adsabs.harvard.edu/abs/1998MNRAS.294..291M}
}

@ARTICLE{2025Baldi_,
       author = {{Bottazzi Baldi}, Bartolomeo and others},
        title = "{The impact of our peculiar motion on primordial non-Gaussianity measurements using the LIGER4GAL framework}",
      journal = {arXiv e-prints},
         year = 2025,
        month = dec,
          eid = {arXiv:2512.03824},
        pages = {arXiv:2512.03824},
          doi = {10.48550/arXiv.2512.03824},
archivePrefix = {arXiv},
       eprint = {2512.03824},
 primaryClass = {astro-ph.CO},
       adsurl = {https://ui.adsabs.harvard.edu/abs/2025arXiv251203824B}
}

@ARTICLE{2018Bonvin&Fleury,
       author = {{Bonvin}, Camille and {Fleury}, Pierre},
        title = "{Testing the equivalence principle on cosmological scales}",
      journal = {\jcap},
         year = 2018,
        month = may,
       volume = {2018},
       number = {5},
          eid = {061},
        pages = {061},
          doi = {10.1088/1475-7516/2018/05/061},
archivePrefix = {arXiv},
       eprint = {1803.02771},
 primaryClass = {astro-ph.CO},
       adsurl = {https://ui.adsabs.harvard.edu/abs/2018JCAP...05..061B}
}


\end{document}